\documentstyle[12pt]{article}

\catcode`\@=11
\long\def\@makefntext#1{
\protect\noindent \hbox to 3.2pt {\hskip-.9pt
$^{{\ninerm\@thefnmark}}$\hfil}#1\hfill}  

\def\@makefnmark{\hbox to 0pt{$^{\@thefnmark}$\hss}}  

\def\ps@myheadings{\let\@mkboth\@gobbletwo
\def\@oddhead{\hbox{}
\rightmark\hfil\ninerm\thepage}
\def\@oddfoot{}\def\@evenhead{\ninerm\thepage\hfil
\leftmark\hbox{}}\def\@evenfoot{}
\def\sectionmark##1{}\def\subsectionmark##1{}}

\newcounter{sectionc}\newcounter{subsectionc}\newcounter{subsubsectionc}
\renewcommand{\section}[1]
{\vspace{0.6cm}\addtocounter{sectionc}{1}
\setcounter{subsectionc}{0}\setcounter{subsubsectionc}{0}\noindent
           {\large\bf\thesectionc #1}\par\vspace{0.4cm}}
\renewcommand{\subsection}[1]
{\vspace{0.6cm}\addtocounter{subsectionc}{1}
     \setcounter{subsubsectionc}{0}\noindent
     {\it\thesectionc.\thesubsectionc. #1}\par\vspace{0.4cm}}
\renewcommand{\subsubsection}[1]
             {\vspace{0.6cm}\addtocounter{subsubsectionc}{1}
     \noindent {\rm\thesectionc.\thesubsectionc.\thesubsubsectionc.
     #1}\par\vspace{0.4cm}}

\newcounter{appendixc}
\newcounter{subappendixc}[appendixc]
\newcounter{subsubappendixc}[subappendixc]

\renewcommand{\appendix}[1] {\vspace{0.6cm}
        \refstepcounter{appendixc}
        \setcounter{figure}{0}
        \setcounter{table}{0}
        \setcounter{equation}{0}
        \renewcommand{\thefigure}{\Alph{appendixc}.\arabic{figure}}
        \renewcommand{\thetable}{\Alph{appendixc}.\arabic{table}}
        \renewcommand{\theappendixc}{\Alph{appendixc}}
        \renewcommand{\theequation}{\Alph{appendixc}.\arabic{equation}}
        \noindent{\bf Appendix \theappendixc #1}\par\vspace{0.4cm}}

\def\abstracts#1{{

\centering{\begin{minipage}{26pc}\tenrm\baselineskip=12pt\noindent
     \parindent=0pt #1
     \end{minipage}}\par}}

\renewenvironment{thebibliography}[1]
     {\begin{list}{[\arabic{enumi}]}
     {\usecounter{enumi}\setlength{\parsep}{0pt}
\setlength{\leftmargin 0.6cm}{\rightmargin 0pt}
      \setlength{\itemsep}{0pt} \settowidth
     {\labelwidth}{#1.}\sloppy}}{\end{list}}

\newcounter{itemlistc}
\newcounter{romanlistc}
\newcounter{alphlistc}
\newcounter{arabiclistc}

\newcommand{\fcaption}[1]{
        \refstepcounter{figure}
        \setbox\@tempboxa = \hbox{\tenrm Fig.~\thefigure. #1}
        \ifdim \wd\@tempboxa > 6in
           {\begin{center}
        \parbox{6in}{\tenrm\baselineskip=12pt Fig.~\thefigure. #1}
            \end{center}}
        \else
             {\begin{center}
             {\tenrm Fig.~\thefigure. #1}
              \end{center}}
        \fi}

\newcommand{\tcaption}[1]{
        \refstepcounter{table}
        \setbox\@tempboxa = \hbox{\tenrm Table~\thetable. #1}
        \ifdim \wd\@tempboxa > 6in
           {\begin{center}
        \parbox{6in}{\tenrm\baselineskip=12pt Table~\thetable. #1}
            \end{center}}
        \else
             {\begin{center}
             {\tenrm Table~\thetable. #1}
              \end{center}}
        \fi}

\def\@citex[#1]#2{\if@filesw\immediate\write\@auxout
     {\string\citation{#2}}\fi
\def\@citea{}\@cite{\@for\@citeb:=#2\do
     {\@citea\def\@citea{,}\@ifundefined
     {b@\@citeb}{{\bf ?}\@warning
     {Citation `\@citeb' on page \thepage \space undefined}}
     {\csname b@\@citeb\endcsname}}}{#1}}

\def\fnt#1#2{\footnotetext{\kern-.3em
     {$^{\mbox{\sevenrm #1}}$}{#2}}}

 1
 1
 1

\font\tenbf=cmbx10
\font\tenrm=cmr10

\font\ninerm=cmr9

\input psfig.tex
\begin{document}
\hfill RUB-TPII-04/96

\centerline{\tenbf END-POINT BEHAVIOR OF EXCLUSIVE PROCESSES: THE}
\centerline{\tenbf TWILIGHT REGIME OF PERTURBATIVE QCD
\footnote{Invited plenary talk presented at {\it HADRON STRUCTURE '96.}
          High Energy Interactions: Theory and Experiments, Star\'a
          Lesn\'a, High Tatra, Slovac Republic, February 12-17, 1996}
          }
\baselineskip=3pt
\vspace{0.7cm}
\centerline{\tenbf N.\ G.\ STEFANIS}
\baselineskip=13pt
\centerline{\tenrm Institut f\"ur Theoretische Physik~II,
                   Ruhr-Universit\"at Bochum}
\baselineskip=13pt
\centerline{\tenrm D-44780 Bochum, Germany}
\vspace{0.25cm}
\abstracts{\ninerm
A selected set of topics along the borderline between perturbative
and nonperturbative QCD in exclusive reactions are studied. Specific
problems, related to different mechanisms of momentum transfer to an
intact hadron, are discussed. Calculations of the space-like form
factors of the pion and the nucleon are reviewed within a convolution
scheme of short-distance (hard) and large-distance (soft)
contributions which takes into account soft gluon emission and the
intrinsic transverse hadron size. The failure of this scheme to
reproduce the existing experimental data signals sizeable
higher-order perturbative corrections (a K-factor of order two)
and/or higher-twist contributions.}

\vspace{0.6cm}
\rm\baselineskip=12pt
\textheight 18.8 truecm
\tenrm
\section{$\;\;$ Introduction}
\label{sec:Intro}
\noindent
This article presents an overwiew of exclusive processes, focusing
on their end-point behavior. To set the stage, we discuss and review
problems relating to the (momentum) scales involved in form factor
calculations: scale locality, infrared (IR) safety, gluonic radiative
corrections, and the role of hadronic size effects. These issues are
more precisely described in terms of the essential mechanisms of
momentum transfer to an intact hadron. We then use detailed
calculations to investigate how these effects influence the
predictions for
$F_{\pi }(Q^{2})$, $G_{M}^{p}(Q^{2})$, and $G_{M}^{n}(Q^{2})$
relative to existing data.

The application of perturbative Quantum ChromoDynamics (pQCD) to
inclusive processes has been very successful and predictive and there
is now ample experimental verification for a variety of reactions.
In contrast, exclusive processes, though of basic importance for a
deeper understanding of confinement, are yet not so rigorously
established.
In order that pQCD becomes applicable at the {\tensl amplitude
level}, a short-distance part of the strong-interaction amplitude
has to be isolated. This is then amenable to perturbative analysis
within a hard-scattering scheme.
Beyond this, however, one has to use additional (unsettled)
nonperturbative methods to model the hadron wave functions which
encode bound-state features. Since hadron wave functions appear in
{\tensl integrated} quantities they are not directly accessible to
experiment.

Once factorization of regimes has been accomplished, renormalization
group (RG) techniques can be employed to calculate the evolution
behavior of the factorized parts. The logarithmic scaling violations
are found to be controlled by the same nonsinglet anomalous
dimensions as in deep-inelastic scattering~\cite{LB80,Pes79}.

\section{$\;\;$ Factorization in exclusive processes}
\label{sec:Fact}
\noindent
Factorization theorems are of central importance in quantum field
theory. The basic idea is that one can separate high-momentum
from low-momentum dependence in a multiplicative way. For example,
proving that ultraviolet (UV) divergences occuring in Feynman graphs
can be absorbed into multiplicative renormalization factors (infinite
constants) is instrumental in establishing renormalizability of the
theory.
The technical difficulty is to prove factorization of a particular
QCD process to {\tensl all-orders} in the coupling constant going
beyond leading logarithms~\cite{CSS89}. These difficulties derive
from the fact that in QCD a new type of IR-divergence is encountered,
the {\tensl collinear} divergence, and that in higher orders the
self-coupling of gluons becomes important in the exponentiation of
IR-divergences.

The realization of factorization when applying to elastic form
factors can be written in the form of a convolution of a
hard-scattering amplitude (dubbed $T_{H}$) describing the
short-distance quark-gluon interactions, and two soft wave functions
corresponding to the incoming and outcoming hadron~\cite{LB80,ER80}.
Generically,
\begin{equation}
  F(Q^{2})
=
  \Phi ^{out}(m/\mu )
  \otimes
  T_{H}(\mu /Q )
  \otimes
  \Phi ^ {in}(m/\mu ) \; ,
\label{eq:F(Q^2)}
\end{equation}
where $m$ sets the typical virtuality in the soft parts and $Q$
is the (external) scale characteristic of the hard (parton)
subprocesses.
The matching scale $\mu$ at which factorization has been performed
is arbitrary and, assuming that $\mu \gg m$, it can be safely
identified with the renormalization scale -- unavoidable in any
perturbative calculation -- by virtue of the RG equations.
In this way, $F$ can be rewritten as a function only of the
coupling constant operative at that scale.

As long as {\tensl scale locality} is preserved, i.e., the variation
of the effective coupling constant with $\mu$ is governed by the
same momentum scale, and the limit $m \to 0$ is finite,
Eq.~(\ref{eq:F(Q^2)}) is valid because {\tensl intrusions} from the
hard into the soft regime are prohibited. This means that $T_{H}$ is
insensitive to long-distance interactions, i.e., it is
{\tensl IR safe}.
All IR-sensitivity resides in the hadron distribution amplitudes
$\Phi ^{in(out)}$ which are independent of large momentum scales and
may depend on the external scale $Q^{2}$ only through RG-evolution.
Both the subtraction procedure of UV poles in the soft parts
and the cancellation of IR divergences in the hard part are not
uniquely fixed. Nevertheless, they have to ensure that the asymptotic
behavior of $F$ is IR-insensitive and governed by the leading
anomalous dimensions associated with vertex and quark self-energy
corrections.

Adopting a factorization scheme, the initial (final) state of the
hadron has a certain probability distribution for finding its
valence quarks carrying longitudinal momentum fractions
$0\leq x_{i}=k_{i}^{+}/P^{+}\leq 1$ in a $P^{3}\to \infty$ frame.
Apart from the slow perturbative $Q^{2}$-evolution, this randomness
depends only on the uncalculable confinement dynamics and not on the
specific hard-scattering collision, i.e., it is {\tensl universal}.
Hence
\begin{equation}
  \Phi (x_{i},\mu ^{2})
\equiv
  \left(
        \ln \frac{\mu ^{2}}{\Lambda _{QCD}^{2}}
  \right)^{-c\,\gamma _{F}/\beta}
  \int_{0}^{\mu ^{2}} \prod_{i=1}^{N} [d^{2}{\vec k}_{\perp}^{i}]
  \psi (x_{i},{\vec k}_{\perp}^{(i)}) \; ,
\label{eq:Phi-def)}
\end{equation}
where $N=2$, $c=1$ for the meson and $N=3$, $c=2/3$ for the nucleon,
respectively, and $\gamma _{F}$ is the anomalous dimension associated
with quark self-energy.

To solve the evolution equation, $\Phi ^{(H)}$ for hadron $H$ has to
be expressed as an orthogonal expansion in terms of appropriate
hypergeometric functions which constitute an eigenfunction basis of
the gluon-exchange potential, i.e.,
\begin{equation}
  \Phi ^{(H)}(x_{i},Q^{2})
=
  \Phi _{as}^{(H)}(x_{i})
  \sum_{n=0}^{\infty} B_{n}^{(H)}(\mu ^{2})
  \tilde{\Phi} _{n}^{(H)}(x_{i})
  \exp
      \left\{
             \int_{\mu ^{2}}^{Q^{2}}
             \frac{d\bar{\mu}}{\bar{\mu}}
             \gamma _{F}(g(\bar{\mu} ^{2}))
      \right\} ,
\label{eq:Phieigen}
\end{equation}
where $\Phi _{as}^{(H)}$ is the asymptotic amplitude (see below) over
fractional momenta proportional to the weight $w(x_{i})$ of the
orthogonal basis, and $\tilde{\Phi} _{n}^{(H)}$ denote the
corresponding eigenfunctions.
The coefficients $B_{n}^{(H)}$ of this expansion are associated with
matrix elements of composite lowest-twist operators with definite
anomalous dimensions taken between the vacuum and the external
hadron.
In the meson case (leading twist=2), the eigenfunctions of the
diagonalized evolution equation~\cite{LB80} are the Gegenbauer
polynomials $C_{n}^{3/2}$~\cite{Erd53},
which correspond to conformal operators~\cite{ER80a} with associated
anomalous dimensions given by
\begin{equation}
  \gamma _{n}^{(\pi )}
=
  \frac{C_{F}}{\beta}
\left[
      1 + 4\sum_{n=0}^{n+1} \frac{1}{k} - \frac{2}{(n+1)(n+2)}
\right]
\geq 0 \quad\quad (n\;\; {\rm even}) \; .
\label{eq:gamma_pi}
\end{equation}
Introducing the relative coordinate $\xi =x_{1} -x_{2}$,
orthogonality with respect to the weight $w(\xi )=(1-\xi ^{2})$
yields expansion coefficients going like
\begin{equation}
  B_{n}^{(\pi )}
       \left(
             \ln \frac{Q^{2}}{\Lambda _{QCD}^{2}}
       \right)^{-\gamma _{n}}
=
  \frac{2(2n+3)}{(2+n)(1+n)}\int_{-1}^{1}d\xi\,
  C_{n}^{3/2}(\xi )\; \Phi ^{(\pi )}\!(\xi ,Q^{2}) ,
\label{eq:expcoef-pi}
\end{equation}
meaning that for higher orders they decrease like $1/n^{2}$,
provided
$
 \Phi ^{(\pi )}(x_{i}, \mu )\leq K x_{i}^{\epsilon}
$
as $x_{i} \to 0$ for some $\epsilon > 0$~\cite{LB80}.

In the nucleon case (leading twist=3), an orthogonal normalized
basis of the evolution kernel is provided by linear combinations of
Appell polynomials which depend on two variables~\cite{LB80,Ste94}.
Here orthogonality alone is insufficient to fix the eigenfunctions
uniquely. It was first shown in~\cite{ELBA93} that the expansion
coefficients $B_{n}$ become analytically tractable up to any
desired polynomial order $M\geq i+j$ in terms of {\tensl strict}
moments
\begin{equation}
  \Phi _{N}^{(i0j)}(\mu ^{2})
=
  \int_{0}^{1}[dx]\,x_{1}^{i}\,x_{2}^{0}\,x_{3}^{j}\,
  \Phi _{N}(x_{k}, \mu ^{2})
\label{eq:strictmom}
\end{equation}
of the mixed-symmetry nucleon distribution amplitude~\cite{CZ84}
$\Phi _{N}$:
\begin{equation}
  \frac{B_ {n}^{(N)} (Q^{2})}{\sqrt{N_{n}}} =
  \frac{\sqrt{N_{n}}}{120}\,
  B_ {n}^{(N)} (\mu ^{2})
                 \Biggl[
                        \frac {\ln (Q^{2}/\Lambda _{QCD}^{2})}
                        {\ln (\mu ^{2}/\Lambda _{QCD}^{2})}
                 \Biggr]^{-\gamma _{n}}
  \sum_{i,j=0}^{\infty}a_{ij}^{n}\
  \Phi _{N}^{(i0j)}(\mu ^{2}) \; .
\label{eq:Bviamom}
\end{equation}
The projection coefficients $a_{ij}^{n}$, the anomalous dimensions of
trilinear twist-three quark operators
\begin{equation}
 \gamma _{n}^{(N)}
=
 \frac{1}{\beta}
 \left(
       \frac{3}{2}C_{F} + 2\eta _{n}C_{B}
 \right)
\label{eq:gamma_N}
\end{equation}
(where $\eta _{n}$ are the zeros of the characteristic polynomial
that diagonalizes the evolution equation), and the normalization
constants $N_{n}$ up to order $M=4$ are tabulated in~\cite{Ste94}.

The moment values -- calculated, for instance, via QCD sum
rules~\cite{CZ84,COZ89} -- provide only {\tensl local} constraints
that are not sufficient for the distribution amplitude to be
reconstructed in an unambigous way~\cite{Ste89}. One can always add
some oscillating function which vanishes at the points fixed by the
local constraints but which contributes outside. Thus one has to
impose {\tensl global} constraints as well, for shaping the
distribution amplitude as a {\tensl whole}. Such constraints have
been successfully used in~Refs. \cite{Ste94} - \cite{Ste95} to
ensure dominance of the lowest-order contributions and minimize the
influence of disregarded higher-order terms. In~\cite{BS93} a
complete set of nucleon distribution amplitudes was determined
which satisfy existing QCD sum rules~\cite{COZ89,KS87} with
comparable degree of accuracy while avoiding unphysical oscillations.
These solutions organize themselves across a ``fiducial orbit''
that is characterized by a scaling relation between the form-factor
ratio $|G_{N}^{n}|/G_{M}^{p}$ and the expansion coefficient
$B_{4}$ (cf.~Eq.~(\ref{eq:Bviamom})).

Hadron distribution amplitudes derived this way from QCD sum rules
show an asymmetric balance of longitudinal momentum fractions of
valence quarks. Convoluted with the corresponding hard-scattering
amplitudes they lead to form factors which have the right magnitude
and QCD-evolution behavior~\cite{Ste94}.
On the other hand, it was pointed out in~\cite{ILS89,Rad91} that
asymmetric distribution amplitudes enhance the contributions of
endpoint regions and that, extracting these regions, the leading
perturbative contribution to the form factor is reduced to a small
fraction. This depletion of the form factor indicates sensitivity
to the gluon offshellness in the end-point region, presumed to be
large. Hence the perturbative treatment turns out to be inconsistent,
meaning that uncalculated higher-order terms may be important, even
leading.
To reinstate the validity of pQCD, end-point contributions have
clearly to be suppressed. To this end, a modified convolution
scheme~\cite{LS92} -- still within the hard-scattering picture --
will be discussed below which incorporates Sudakov suppression due
to radiative gluon corrections.

\section{$\;\;$ Momentum transfer mechanisms}
\label{sec:Mechanisms}
\begin{figure}
\begin{picture}(0,40)
  \put(25,-60){\psboxscaled{600}{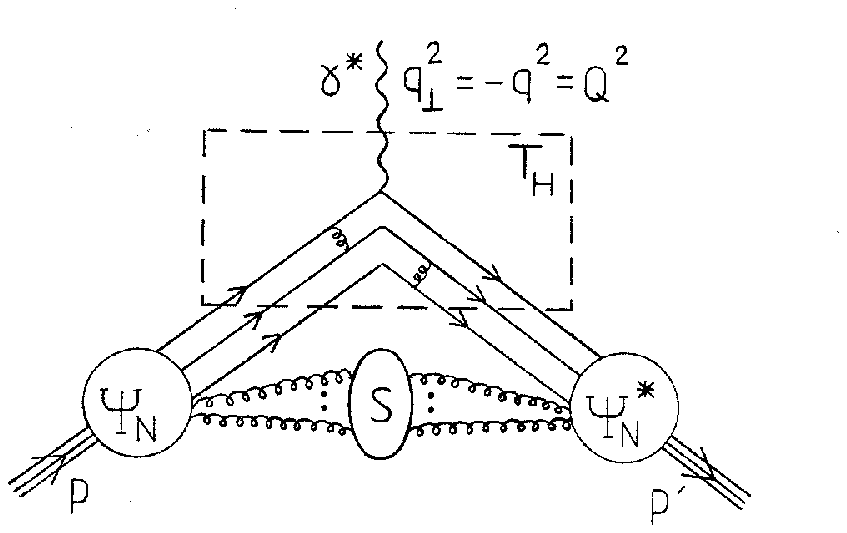}}
\end{picture}
\begin{picture}(0,40)
  \put(195,-48){\psboxscaled{600}{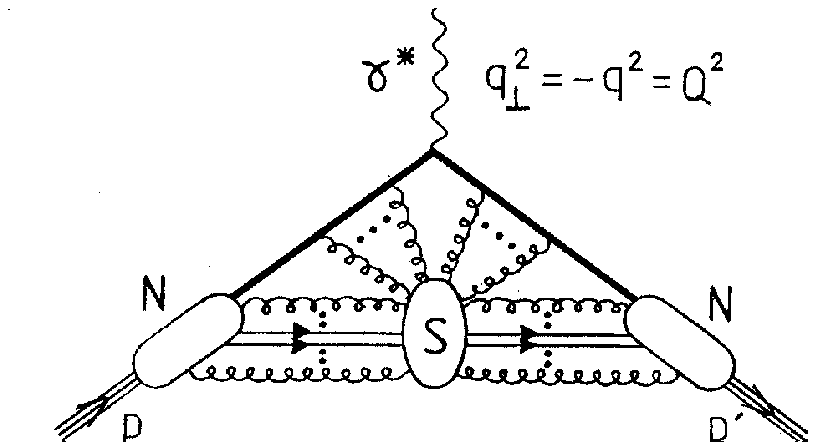}}
\end{picture}
\vspace{1.7 true cm}
\caption[fig:mech]
{\tenrm  Mechanisms for momentum transfer during elastic scattering.
         The lhs shows hard-gluon exchange within pQCD.
         The blob $S$ containing soft gluon lines (and an analogous
         one with soft quark-antiquark lines not shown here) spoils
         factorization but is power-suppressed, i.e., non-leading.
         The (rhs) shows the Feynman mechanism using for purposes
         of illustration quark and gluon lines. The leading quark
         is denoted by a heavy line, while all other lines represent
         wee quarks and soft gluons.}
\label{fig:momtranmechs}
\end{figure}

\noindent
There are basically two schemes for describing the transfer of a
large external momentum $Q$ to an intact hadron during elastic
scattering: hard-gluon exchange and the Feynman mechanism. We begin
with the first one which is tightly connected to pQCD and relies on
the factorization theorem. Following this rationale, the struck quark
connects to the other valence quarks via highly off-shell gluon
propagators, meaning that the transverse interquark distances are
rather small, {\tensl viz.} of order $1/Q$ and that {\tensl all}
partons share comparable fractions of longitudinal momentum.
Thus $T_{H}$ can be reliably computed as a power series in the
running coupling constant
$\alpha _{s}$~\cite{LB80,ER80a}
(see the lhs of Fig.~\ref{fig:momtranmechs}).

In the asymptotic limit $Q^{2}\to \infty$, only the $n=0$ term in
Eq.~(\ref{eq:Phieigen}) survives so that the RG-asymptote of
the distribution amplitude reads
($\gamma _{0}<\gamma _{n}$ for all $n > 0$)
\begin{equation}
  \Phi _{as}^{(H)}(Q^{2} \to \infty )
=
  B_{0}^{(H)}w(x_{i})
  \lim_{Q^{2}\to\infty}
  \left(
        \ln \frac{Q^{2}}{\Lambda _{QCD}}
  \right)^{-C/\beta}
\label{eq:Phiasy}
\end{equation}
with $C=C_{F}$ (pion) and $C=3C_{F}/2 - 2C_{B}$ (nucleon),
where $B_{0}^{(H)}$ is the hadron wave function at the origin of
coordinate space and the limit of the logarithm amounts to the
wave-function renormalization factor $Z_{2}$.
From these distribution amplitudes one infers that asymptotically
the most likely configurations are those in which the valence quarks
share longitudinal momentum in a uniform way, i.e., $x_{i}=1/2$
for the pion and $x_{i}=1/3$ for the nucleon. Within this scheme,
when confinement sets in, a quark is not able to venture too far
from the antiquark (in the pion) or the other two valence quarks
(in the nucleon). This poses constraints on the offshellness of the
involved propagators, typified by
$
 x_{i}x_{j}^{\prime}
>
 \left(
       \Lambda _{QCD}^{2}/Q^{2}
 \right)
 \exp{\left(4\pi /\beta \alpha _{s}\right)}
$.
The problem then is that in the end-point regions
$x_{i},x_{j}^{\prime} \to 0,1$
the gluons become nearly on-shell (i.e., real) and the hard-gluon
exchange mechanism becomes unreliable.
This is also reflected in the behavior of form factors.
As a rule, narrow distribution amplitudes yield for reasonable
values of $\alpha _{s}$ results which are unrealistically low to
be consistent with the data. Obviously, {\tensl broad} distribution
amplitudes are required to match the data.

One method~\cite{CZ84} extracts distribution amplitude moments from
QCD sum rules using {\tensl local} vacuum condensates. In the pion
case, the large values obtained this way for the moments can only be
realized by a ``double-humped'' distribution amplitude of the form
$
 \phi _{CZ}^{(\pi )}(x)
=
30 f_{\pi}x(1-x)(1-2x)
$
which is end-point dominated, whereas the central region
($x_{i}=1/2$) is depleted. This distribution amplitude yields a pion
form factor in much better agreement with data but at the expense
that it accumulates its main contributions from the end-point region
where a perturbative treatment is less reliable.

The approach taken in~\cite{MR89} is conceptually quite different.
To avoid the inherent deficiencies of moment inversion (see for
criticism ~\cite{Rad91}), the pion distribution amplitude was
computed directly from QCD sum rules via dispersion relations.
This approach makes use of {\tensl nonlocal} vacuum condensates
which afford for the finite average virtuality of the vacuum quarks.
Parametrizing the nonlocal quark condensate by a Gaussian, a model
distribution amplitude was obtained which gives lowest moments
$<\xi ^{n}>$ with $n=2,4,6$ close to those of
$
 \Phi _{as}^{(\pi )}
=
 f_{\pi}6x(1-x)
$,
but which has a significantly wider shape and no dip in the central
region.

Concerning the nucleon the situation is technically more complicated,
though, perhaps, less prone to criticism. Up to now, all derived
model distribution amplitudes (see, e.g.,~\cite{Ste94,Ste95}) rely
on moment-inversion techniques within rather large uncertainty
intervals. However, a consistent pattern has
emerged~\cite{Sch89,BS93} which seems to encapsulate the main
characteristics of the true nucleon distribution amplitude.

Let us consider now the other basic mechanism for elastic scattering
due to Feynman~\cite{Fey72}.
In this scheme, almost all of the hadron's momentum is carried off
by a single parton, the others being ``wee''. This picture is
consistent with a configuration in which only the struck quark is
within an impact distance $1/Q$ of the electron while all other
partons have rather random positions in the transverse direction,
building a soft ``cloud'' with transverse size $\gg 1/Q$.
Once the elastic scattering has happened, rearrangements are
necessary to change quarks and gluons into hadrons.
This conversion procedure (visualized on the rhs of
Fig.~\ref{fig:momtranmechs}) is controlled by the overlap of the
initial and final state wave functions and cannot be computed
within pQCD.

%
\input psbox.tex
%
For the remainder of this report we will consider only calculations
which are based on the hard-scattering picture.

\section{$\;\;$ Modified Convolution Scheme}
\label{sec:MCS}
\noindent
In deriving Eq.~(\ref{eq:F(Q^2)}), we tacitly assumed that the
$k_{\perp}$-dependence of the quark and gluon propagators in $T_{H}$
can be ignored. This is tantamount to factorizing the
$k_{\perp}$-dependence into the distribution amplitudes which are
the wave functions integrated over $k_{\perp}$ up to the
factorization scale. Then, in the limit $Q^{2}\to\infty$, the only
gluon radiative corrections remaining uncancelled are those giving
rise to wave-function renormalization. Hoewever, in the end-point
region the parton transverse momenta in $T_{H}$ cannot be {\tensl a
priori} ignored since, say, for the pion,
$
 \left(\vec{k}^{}_{\perp i} + \vec{k}_{\perp j}^{\prime}\right)^{2}
\gg
 x^{}_{i}x_{j}^{\prime}Q^{2}
$.
As a result, the transverse distance between the quark and the
antiquark becomes large compared to $1/Q$ and the corresponding
gluon line is no more part of the hard-scattering process but
should be counted to its soft part. In other words, the hard-gluon
exchange mechanism should be replaced by that of Feynman.

The physical basis of the modified convolution scheme
(MCS)~\cite{LS92} is to dissect the process in such a way, so that
for transverse distances large compared to $1/Q$ (playground of the
hard-scattering mechanism) but still small relative to the true
confinement regime -- characterized by $1/\Lambda _{QCD}$ -- the
hadron wave function is modified to exhibit the effect of Sudakov
enhancements explicitly up to the transverse scale retained in
$T_{H}$. Going over to transverse configuration space, the modified
wave function reads
\begin{equation}
  \hat{\Psi} ^{(H)}_{(mod)}(x_{i}, 1/\tilde{b}_{i}, Q, \mu _{ren})
=
  {\rm e}^{-S}\hat{\Psi} ^{(H)}(x_{i}, 1/\tilde{b}_{i})
\label{eq:modPsi} \; ,
\end{equation}
where the factor
\begin{equation}
  \exp{(-S)}
=
  \sum_{i=1}^{N}
  \left[
          s(x_{i}, \tilde{b}_{i}, Q)
        +
          \int_{1/\tilde{b}_{i}}^{\mu _{ren}}
          \frac{d\overline{\mu}}{\overline{\mu}}
          \gamma _{q}(g(\overline{\mu}^{2}))
  \right]
+ \quad x_{i} \leftrightarrow x_{i}^{\prime}
\label{eq:Sudexp}
\end{equation}
comprises gluon corrections in terms of the functions
$s(x_{i}, \tilde{b}_{i}, Q)$~\cite{LS92}
and accounts for RG-evolution from the IR-scale $1/\tilde{b_{i}}$ to
the renormalization scale $\mu _{ren}$ via the quark anomalous
dimension (in the axial gauge)
$
 \gamma _{q}(g(\overline{\mu}^{2}))
=
 -\alpha _{s}/\pi \; + \; O(\alpha _{s}^{2})
$.
The explicit expressions for the Sudakov functions are given
in~\cite{BS89}.
The Sudakov exponential factor resums contributions from two-particle
reducible diagrams (giving rise to double logarithms), whereas
two-particle irreducible diagrams (giving rise to single logarithms)
are absorbed into the hard scattering amplitude
$T_{H}$~\cite{BS89}. It can be conceived of as a {\tensl finite}
renormalization factor to the hadron wave function~\cite{Ste95}.
The leading double logarithms derive from those momentum regions
where soft gluons (all four-momentum components small) and collinear
gluons to the external quark lines overlap. These contributions are
numerically dominated by the term
\begin{equation}
 \exp
     \left\{
            - \frac{2C_{F}}{\beta}
              \ln \frac{\xi _{i}Q}{\sqrt{2}\Lambda _{QCD}}
              \ln\frac{\ln \left(\xi _{i}Q/\sqrt{2}\Lambda _{QCD}
                           \right)
                      }
              {\ln\left(1/\tilde{b}_{i}\Lambda _{QCD}\right)},
     \right\}
\label{eq:doublelog}
\end{equation}
where $\xi _{i}$ is one of the fractions $x_{i}$ or $x_{i}'$,
and $\beta = (33-2n_{f})/3$ is the first-order term of the
Gell-Mann and Low function encountered before.
The single logarithm stems from the running coupling constant and
the double logarithm contains the exponentiated higher-order
corrections -- required by RG -- rendered finite by the
{\tensl inherent} IR-cutofff $1/\tilde{b}_{i}$. This marks the
crucial difference between the MCS and previous approaches dealing
with {\tensl isolated} quarks where such IR-cutoff parameters had
to be introduced as {\tensl external} regulators.

For small transverse distances (or equivalently,
$1/\tilde{b}_{i}\gg \xi _{i}Q$),
gluonic radiative corrections are treated as being part of $T_{H}$
and are excluded from the Sudakov form factor. Consequently, for
$\xi _{i}\le\sqrt{2}/\tilde{b}_{i}Q$ the Sudakov functions
$s(\xi _{i},\tilde{b}_{i},Q)$ are set equal to zero.
On the other hand, as $\tilde{b}_{i}$ increases $e^{-S}$ decreases,
reaching zero at $\tilde{b}_{i}\Lambda _{QCD}=1$.
In the pion case, there is only one transverse scale, notably, the
quark-antiquark separation $b$, and suppression is automatically
accomplished.
Indeed, when it happens that one Sudakov function
$s(\xi ,\tilde{b}_{i},Q)=0$ (or equivalently that the corresponding
exponential is set equal to unity) the other (negative) Sudakov
function in the exponent, $s(1-\xi ,b,Q)$, diverges, thus providing
sufficient suppression.

Our approach~\cite{BJKBS95} to the choice of the appropriate
IR-cutoff in calculating observables with several transverse
momentum scales involved -- like nucleon form factors -- is to
postulate that long wave-length (compared to the typical interquark
separations) gluons ``see'' the nucleon as a {\tensl whole}, i.e.,
in a color-singlet state and cannot resolve its color structure.
This is technically implemented by setting
$
  \tilde{b}\equiv {\rm max}\{b_{1},b_{2},b_{3}\}
=
  \tilde{b}_{1}=\tilde{b}_{2}=\tilde{b}_{3}
$
(``MAX'' prescription).
As a result, (i) logarithmic $\alpha _{s}$-singularities in the
end-point region are {\tensl screened} by the exponentially
decreasing Sudakov factors, (ii) the form-factor integrands are
IR-safe for all possible kinematic configurations and receive
perturbative contributions which {\tensl saturate}, i.e., which are
rather insensitive to distances of order $1/\Lambda _{QCD}$.
It was outlined in~\cite{BJKBS95,Ste95} that other choices may lead
to uncompensated $\alpha _{s}$-singularities.
In what follows we present those aspects of the calculation which
differ from the standard case. We also present theoretical
predictions in comparison with available experimental data. A more
detailed level of description is given in~\cite{BJKBS95,Ste95}.

\begin{figure}[h]
\[
  \psfig{figure=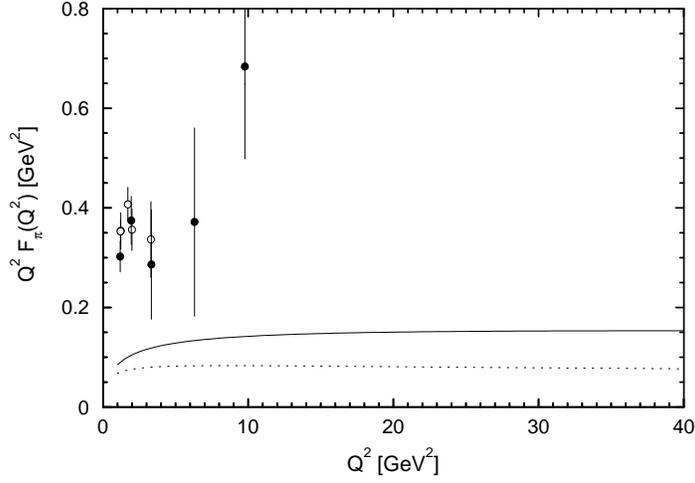,%
        bbllx=2.5cm,bblly=14cm,bburx=17.5cm,bbury=25cm,%
        width=8cm}
\]
\vspace{-1cm}
\caption[fig:pion]
{\tenrm  Spacelike pion form factor comprising Sudakov corrections
         and the intrinsic transverse size of the pion wave
         function~\cite{JK93} in comparison with experimental
         data.
         The curves correspond to the Chernyak-Zhitnisky
         model~\cite{CZ84} (solid line) and the asymptotic wave
         function (dotted line). The data are from~\cite{Beb76}.}
\label{fig:pionffsudkperp}
\end{figure}

\begin{figure}
\unitlength 1mm
\begin{picture}(0,40)
  \put( 0,-100){\psboxscaled{500}{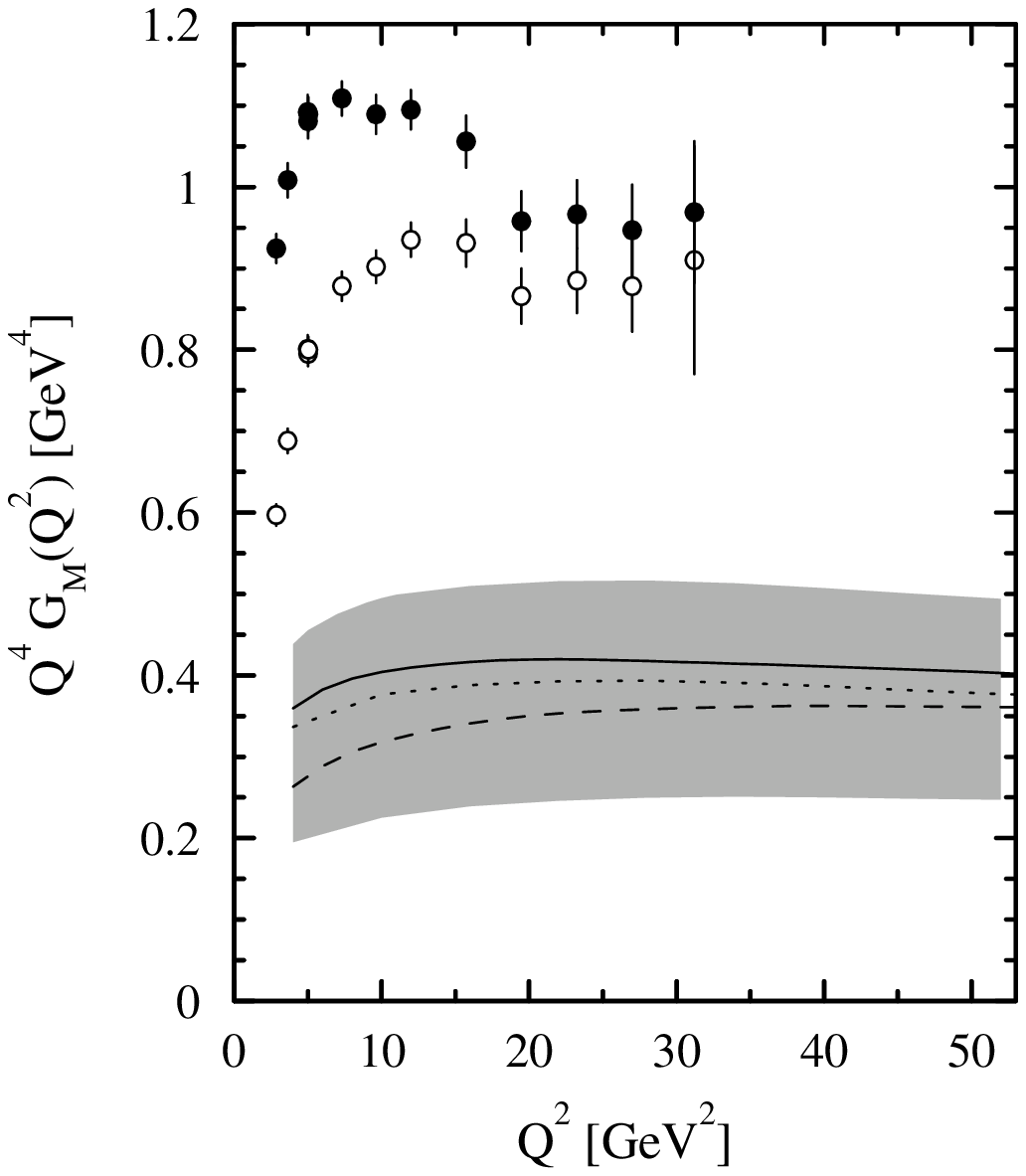}}
\end{picture}
\begin{picture}(0,40)
  \put(60,-100){\psboxscaled{500}{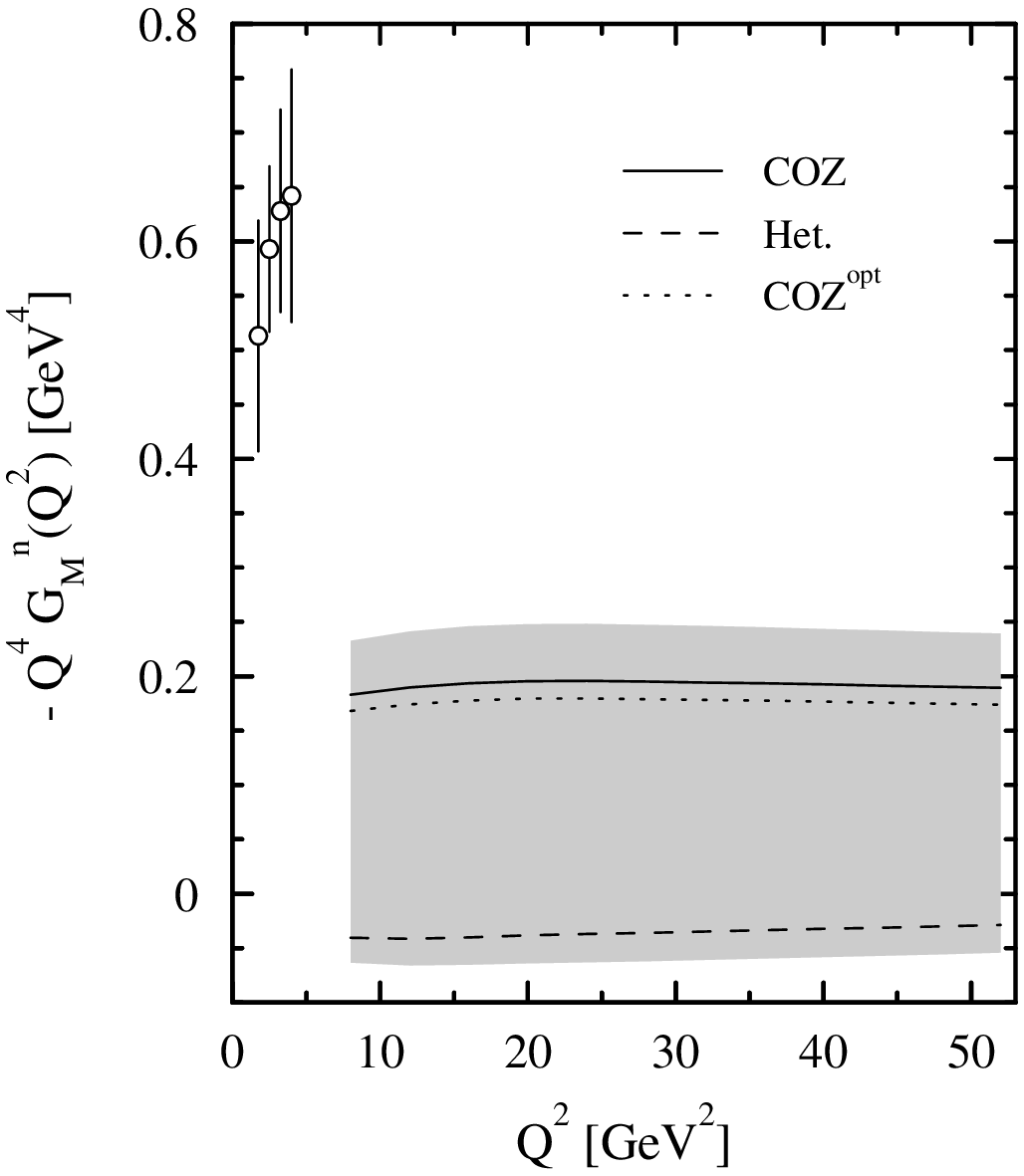}}
\end{picture}
\vspace{1.7 true cm}
\caption[fig:nucleon]
        {\tenrm
         Spacelike magnetic form factors of the nucleon:
         proton (lhs) and neutron (rhs) ($\Lambda _{QCD}=180$~MeV).
         The open circles indicate $F_{1}^{p}$ data~\cite{Sil93}.
         The calculated curves have been obtained in~\cite{BJKBS95}
         with the ``MAX'' prescription, and including QCD-evolution
         and the intrinsic $k_{\perp}$-dependence of the nucleon
         wave function (the latter normalized to unity). The shaded
         area contains the predictions derived for the set of
         nucleon distribution amplitudes determined in~\cite{BS93}
         via QCD sum rules.}
\label{fig:nuclffsudkperp}
\end{figure}

All told, the nucleon form factor recast in the transverse
configuration space reads
\begin{eqnarray}
  G_{M}(Q^{2})
\!\!\! & = & \!\!\!
  \frac{16}{3}
  \int_{0}^{1}[dx][dx']
  \int_{}^{}\frac{d{}^{2}b_{1}}{(4\pi )^{2}}
            \frac{d{}^{2}b_{2}}{(4\pi )^{2}}
  \sum_{j}^{}\, \hat{T}_{j}(x,x',\vec{b},Q,\mu )
  \hat{Y}_{j}(x,x',\vec{b},\mu )
\nonumber \\
& \times & \!\!\!
  \exp{\left[-S_{j}(\xi _{i}, \tilde{b}_{i}, Q, \mu )\right]} \; ,
\label{eq:G_M(b)}
\end{eqnarray}
where the transverse separation vectors between quarks $1$ and $3$,
$2$ and $3$, and $1$ and $2$ are defined as follows:
$\vec{b}_{1}=\vec{b}_{1}^{\prime}$,
$\vec{b}_{2}=\vec{b}_{2}^{\prime}$,
$\vec{b}_{3}=\vec{b}_{2}-\vec{b}_{1}$.
The modified wave functions are given by
($\overline{x}_{i}\equiv 1-x_{i}$)
\begin{eqnarray}
  \hat{Y}_{1}^{p(n)}  \!\!
& = & \!\!
  \frac{1}{\overline{x}_{1}\overline{x}^{\prime}_{1}}
\Bigl\{
             4 (-2)\hat{\Psi}^{\star\prime}_{123}\hat{\Psi}_{123}
           + 4 (-2)\hat{\Psi}^{\star\prime}_{132}\hat{\Psi}_{132}
           +       \hat{\Psi}^{\star\prime}_{231}\hat{\Psi}_{231}
           +       \hat{\Psi}^{\star\prime}_{321}\hat{\Psi}_{321}
\nonumber \\
\!\! & \pm  &\!\!  2 (1)\left[
                       \hat{\Psi}^{\star\prime}_{231}\hat{\Psi}_{132}
               +       \hat{\Psi}^{\star\prime}_{132}\hat{\Psi}_{231}
               +       \hat{\Psi}^{\star\prime}_{321}\hat{\Psi}_{123}
               +       \hat{\Psi}^{\star\prime}_{123}\hat{\Psi}_{321}
                        \right]
  \Bigr\}\, ,
\label{eq:Y_1^p(n)}
\end{eqnarray}
\begin{eqnarray}
  \hat{Y}_{2}^{p(n)} \!\!
& = & \!\!
  \frac{1(2)}{2\overline{x}_{2}\overline{x}^{\prime}_{1}}
  \left\{
              3 (0)\hat{\Psi}^{\star\prime}_{132}\hat{\Psi}_{132}
         \mp       \hat{\Psi}^{\star\prime}_{231}\hat{\Psi}_{231}
         \mp       \hat{\Psi}^{\star\prime}_{231}\hat{\Psi}_{132}
         \mp       \hat{\Psi}^{\star\prime}_{132}\hat{\Psi}_{231}
  \right\} \mathop{\phantom\sum}\mathop{\phantom\sum}
\nonumber \\
\!\! & \mp & \!\!\!\!
\frac{1}{\overline{x}_{3}\overline{x}^{\prime}_{1}}
    \left\{2 (1)
           \! \left[
                    \hat{\Psi}^{\star\prime}_{321}\hat{\Psi}_{321}
           \! + \!  \hat{\Psi}^{\star\prime}_{321}\hat{\Psi}_{123}
           \! + \!  \hat{\Psi}^{\star\prime}_{123}\hat{\Psi}_{321}
              \right]
\!         \!\pm\!  \hat{\Psi}^{\star\prime}_{123}\hat{\Psi}_{123}
\!  \right\} ,
\label{eq:Y_2^p(n)}
\end{eqnarray}
where the lower signs and the numbers in parentheses refer to the
neutron.
The diagrams of hard-gluon exchanges in the MCS can be conveniently
combined~\cite{LS92} to give

\begin{equation}
  \hat{T}_{1}
=
  \frac{8}{3}\,C_{F}\,\alpha _{s}(t_{11}) \alpha _{s}(t_{12})
  K_{0}
       \left(
             (\overline{x}_{1}\overline{x}_{1}')^{1/2}Qb_{1}
       \right)
  K_{0}\left(
             (x_{2}x_{2}')^{1/2}Qb_{2}
       \right),
\label{eq:FourierT_1}
\end{equation}
\begin{equation}
  \hat{T}_{2}
=
  \frac{8}{3}\,C_{F}\,\alpha _{s}(t_{21}) \alpha _{s}(t_{22})
  K_{0}
       \left(
             (x_{1}x_{1}')^{1/2}Qb_{1}
       \right)
  K_{0}
       \left(
             (x_{2}x_{2}')^{1/2}Qb_{2}
       \right) \; ,
\label{eq:FourierT_2}
\end{equation}
where $K_{0}$ is the modified Bessel function of order 0 (the
Macdonald function).
The arguments of the running coupling constant, $t_{ji}$,
are defined as the maximum scale of either the longitudinal momentum
$\propto Q$ or the inverse transverse separation $\propto 1/b_{i}$,
appearing in the argument of $K_{0}$.
They are associated with the virtualities of the exchanged
gluons, namely,
$
 t_{11}
=
  {\rm max} \left[
                  (\overline{x}_{1}\overline{x}_{1}')^{1/2}Q, 1/b_{1}
            \right],
 t_{21}
=
  {\rm max} \left[
                  (x_{1}x_{1}')^{1/2}Q, 1/b_{1}
            \right]
$,
and
$
 t_{12}
=
  t_{22}
=
  {\rm max} \left[
                  (x_{2}x_{2}')^{1/2}Q, 1/b_{2}
            \right].
$
Note that the Fourier transform of the proton wave function reads
\begin{equation}
  \hat{\Psi}_{123}(x,\vec{b},\mu )
=
  \frac{1}{8\sqrt{N_{c}!}}\,
  f_{N}(\mu )
  \Phi _{123}(x,\mu )
  \hat{\Omega}_{123}(x,\vec{b}) \; ,
\label{eq:FourierPsi}
\end{equation}
where the $k_{\perp}$-dependent part is modeled by~\cite{BJKBS95}
\begin{equation}
  \hat{\Omega}_{123}(x,\vec{b})
=
  (4\pi )^{2}
  \exp
           \left[
                 - \frac{1}{4a^{2}}
                 \Bigl(
                         x_{1}x_{3}b_{1}^{2} + x_{2}x_{3}b_{2}^{2}
                       + x_{1}x_{2}b_{3}^{2}
                 \Bigr)
           \right] \; .
\label{eq:FourierOmega}
\end{equation}
Similar, albeit simplified, expressions are obtained also for the
pion form factor~\cite{JK93}. The theoretical predictions are shown
in Figs.~\ref{fig:pionffsudkperp} and \ref{fig:nuclffsudkperp}.

%
%

\section{$\;\;$ Conclusions}
\label{sec:Concl}
\noindent
The above discussion raises a troubling question. If the
self-consistent calculations within the MCS are insufficient to
describe the existing data, is all hope lost for the applicability
of the pQCD paradigm to exclusive reactions? At one level, no:
higher-order corrections, i.e., a K-factor of order $2$ -- in
principle, also computable within the hard-scattering scheme --
might bridge the gap to the data.
But on a deeper level, the news is sobering.
The results may be interpreted as evidence for the failure of the
hard-scattering mechanism in exclusive reactions at accessible
momentum transfer and this calls for nonperturbative mechanisms.

\baselineskip3mm
\itemsep0pt
\parsep10mm

\end{document}